\RequirePackage{fix-cm}
\documentclass[smallextended,refree]{svjour3}       
\usepackage[utf8]{inputenc}
\smartqed  
\usepackage{amssymb}
\usepackage{latexsym}
\usepackage{graphicx}
\begin{document}

\title{
Fidelity-mediated analysis of the
transverse-field $XY$ chain with the long-range interactions:
Anisotropy-driven multi-criticality
}
\subtitle{}


\author{Yoshihiro Nishiyama
}

\institute{Department of Physics, Faculty of Science,
Okayama University, Okayama 700-8530, Japan}

\date{Received: date / Accepted: date}

\maketitle

\begin{abstract}
The transverse-field $XY$ chain with the long-range interactions
was investigated by means of the exact-diagonalization method.
The algebraic decay rate $\sigma$ of the long-range interaction
is related to the
effective dimensionality $D(\sigma)$,
which governs the criticality of the 
transverse-field-driven phase transition
at $H=H_c$.
According to the 
large-$N$ analysis, 
the phase boundary $H_c(\eta)$
exhibits a reentrant behavior within $2<D < 3.065\dots $, as the $XY$-anisotropy $\eta$
changes.
On the one hand,
as for the $D=(2+1)$ and $(1+1)$
short-range $XY$ magnets,
the singularities have been determined
as 
$H_c(\eta) -H_c(0) \sim |\eta|$ and $ 0 $, respectively,
and the transient behavior around $D \approx 2.5$ remains unclear.
As a preliminary survey, setting $(\sigma,\eta)=(1 ,0.5)$,
we investigate
the phase transition by the agency of the fidelity,
which seems to detect the singularity at $H=H_c$ rather sensitively.
Thereby, under the setting $ \sigma=4/3$ ($D=2.5$),
we cast the fidelity data into the 
crossover-scaling formula with the properly scaled $\eta$,
aiming to determine the multi-criticality around $\eta=0$.
Our result indicates that the multi-criticality is identical to that of
the $D=(2+1)$ magnet,
and $H_c(\eta)$'s linearity might be retained down to $D>2$.

\end{abstract}

\section{\label{section1}Introduction}

The $XY$ chain with the transverse field is attracting much attention 
\cite{Maziero10,Sun14,Karpat14}
in the context of the
quantum information theory
\cite{Luo18,Steane98,Bennett00}.
A key ingredient is that the model covers both the $XX$- and $XY$-symmetric cases,
and a variety of phase transitions occur, as the transverse field
and the $XY$-anisotropy change.
Meanwhile, its extention to the long-range-interaction case
has been made \cite{Adelhardt20}; 
particularly, the limiting cases such as the transverse-field Ising 
and $XXZ$ chains
with the long-range interactions
have been investigated in depth 
\cite{Defenu17,Dutta01,Koffel12,Campana10,Gong16,Fey16,Maghrebi17,Frerot17,Roy18,Puebla19}.
A notable point is that 
the effective dimensionality 
\cite{Defenu17}
\begin{equation}
\label{effective_dimensionality}
D=2/\sigma+1
,
\end{equation}
varies, as the decay rate $\sigma$ of the
long-range interactions obeying the power law, $1/r^{1+\sigma}$
($r$: distance between spins), changes.
The effective dimensionality $D$ governs the 
criticality of the transverse-field-driven phase transition.
In fact,
the $D$ dependence on the criticality for
the classical counterpart \cite{Fisher72,Sak73,Gori17,Angelini14,Joyce66}
has 
been studied extensively.
A peculiarity of the quantum magnet is that 
the long-range interaction 
induces
the asymmetry between the real-space and imaginary-time directions
characterized by the dynamical critical exponent 
$z \ne 1$ \cite{Defenu17}.
Hence, it is significant to access to the ground state 
(infinite imaginary-time system size)
directly 
so as to get rid of the influences caused by the aspect ratio
between the real-space and imaginary-time system sizes.

In this paper,
we investigate the transverse-field $XY$ chain with the long-range interactions
\cite{Adelhardt20}
by means of the exact-diagonalization method,
which enables us to access to the ground state directly.
We devote ourselves to the 
anisotropy-driven multi-criticality
around the $XX$-symmetric point.
So far, the transverse-field-driven criticality 
with the {\em fixed} anisotropy
has been explored in detail \cite{Adelhardt20}.
As for the large-$N$ magnet,
the multi-criticality has been studied
\cite{Wald15},
and
an intriguing reentrant behavior was observed 
in low dimensions,
$2< D < 3.065\dots$.
On the contrary, 
as for the $XY$ magnet.
only the cases of
$D=(2+1)$ \cite{Wald15,Henkel84,Jalal16,Nishiyama19}
and 
$(1+1)$ \cite{Mukherjee11} have been studied,
and the transient behavior in between,
$D \approx 2.5$,
remains unclear.
The aim of this paper is to shed light on
such a fractional-effective-dimensionality regime
by adjusting the algebraic decay rate $\sigma$ of the long-range interactions
carefully
with the aid of the $\sigma \leftrightarrow D$
relation, Eq. (\ref{effective_dimensionality}).

To be specific, 
we present the Hamiltonian for the 
transverse-field $XY$ chain with the long-range interactions
\begin{equation}
\label{Hamiltonian}
{\cal H}= 
-
\frac{1}{{\cal N}} \sum_{i \ne j} J_{ij} 
((1+\eta)S_i^xS_j^x+(1-\eta)S^y_iS^y_j)
-H \sum_{i=1}^N S^z_i 
 .
\end{equation}
Here, the quantum spin-$1/2$ operator ${\mathbf S}_i$ is
placed at each lattice point, $i=1,2,\dots,N$.
The parameters, $H$ and $\eta$, denote the transverse field 
and
$XY$ anisotropy, respectively.
The $XY$ interaction between the $i$-$j$ spins 
decays 
algebraically as
\begin{equation}
J_{ij}=1/\sin(\pi|i-j|/N)^{1+\sigma}
,
\end{equation}
with the decay rate $\sigma$,
and the periodic boundary condition is imposed, {\it i.e.}
${\mathbf S}_{N+1}={\mathbf S}_1$, among the alignment of spins, $\{ {\mathbf S}_{i} \}$.
The denominator ${\cal N}$ stands for the Kac factor
\cite{Homrighausen17,Vanderstraeten18}
\begin{equation}
\label{Kac_factor}
{\cal N}=\frac{2}{N}\sum_{i\ne j}1/\sin(\pi|i-j|/N)^{1+\sigma}
.
\end{equation}

\begin{figure}
\includegraphics{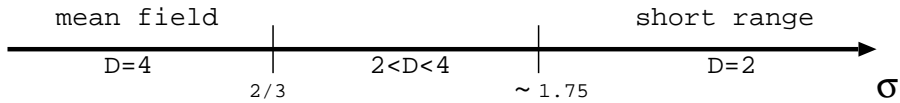}
\caption{
\label{figure1}
The criticality chart 
\cite{Defenu17,Dutta01}
for the transverse-field
$XY$ chain with the long-range interactions 
(\ref{Hamiltonian})
is presented.
The $XY$ case $\eta \ne 0$ is assumed.
As the algebraic decay rate $\sigma$ varies,
there appear a number of universality classes as to the transverse-field-driven phase transition
successively.
For the small decay rate $\sigma \le 2/3  $, the criticality belongs
to the mean-field type with the effective dimensionality $D=4$.
On the contrary,
for $\sigma \gtrsim 1.75$, the ordinary short-range $D=2[=(1+1)]$ universality
comes out.
In the intermediate regime $ 2/3 < \sigma \lesssim 1.75$, 
the criticality changes continuously,
characterized by the effective dimensionality 
ranging within $2< D < 4$.
The upper and lower thresholds,
$\sigma \approx 1.75 $ and $2/3$, are affected by
notorious logarithmic corrections 
\cite{Fey16,Luijten02,Brezin14,Defenu15},
and these regimes lie off the present concern nonetheless.
}
\end{figure}

As shown in Eq. (\ref{effective_dimensionality}),
the effective dimensionality $D=2/\sigma+1$
depends on the decay rate $\sigma$,
which thus governs the
criticality of 
the transverse-field-driven phase transition.
Even quantitatively, the critical exponent for the $D=3$ Ising 
model was pursued \cite{Goll18}
by the $\sigma \leftrightarrow D$ correspondence.
In Fig. \ref{figure1}, we present the criticality chart for the 
$XY$ case ($\eta \ne 1$) \cite{Defenu17,Dutta01}.
For small $\sigma \le 2/3$,
the effective dimensionality $D=4$ 
is realized, 
and the transverse-field-driven criticality belongs to the mean-field type.
On the contrary, for $\sigma \gtrsim 1.75$,
the renormalization-group analysis \cite{Defenu17}
indicates that the long-range interaction becomes
irrelevant, and 
the criticality reduces to that of the short-range magnet;
namely, the $D=2[=(1+1)]$ universality class is realized in this regime.
The threshold $\sigma \approx 1.75$ depends on the internal symmetry group,
namely, either the 
$XX$- ($\eta=0$) or
$XY$-symmetric ($\eta \ne 0$) type
\cite{Defenu17}.
For the intermediate regime
$2/3 < \sigma \lesssim 1.75$,
the effective dimensionality ranges within $2<D<4$.
Rather technically, around both upper and lower thresholds,
there emerge notorious logarithmic corrections
\cite{Fey16,Luijten02,Brezin14,Defenu15},
and these regimes lie off the present concern nonetheless.


\begin{figure}
	\includegraphics[width=8cm]{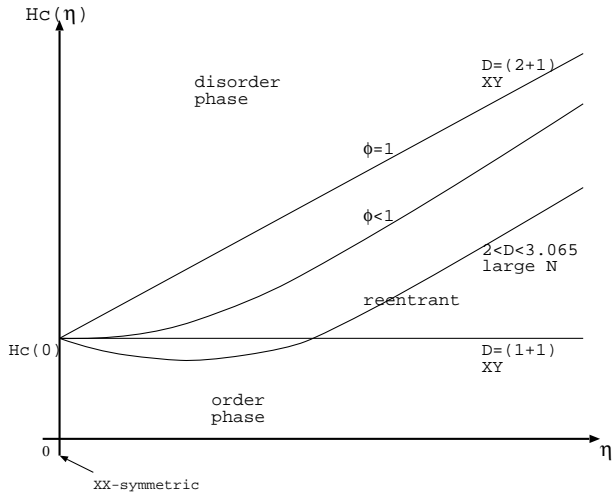}
\caption{
	\label{figure2}
A schematic drawing of the 
transverse-field-driven
	phase boundary
$H_c(\eta)$
with the $XY$ anisotropy $\eta$
for the transverse-field $XY$ model,
is presented.
The short-range 
$D=(2+1)$-dimensional $XY$ magnet
\cite{Wald15,Henkel84,Jalal16,Nishiyama19}
exhibits a linear increase of 
$H_c(\eta)-H_c(0) \sim |\eta|^{1/\phi}$ with the 
crossover exponent $\phi=1$ 
\cite{Riedel69,Pfeuty74},
whereas the $D=(1+1)$ model 
\cite{Mukherjee11}
shows no singularity,
{\it i.e.},
$H_c(\eta) = {\rm const.}$, at all.
It is not clear whether the crossover exponent $\phi$ changes
for the intermediate dimensionality such as $D=2.5$;
in principle, it can be curved convexly, $\phi<1$.
Actually,
as for the large-$N$ magnet \cite{Wald15},
there appear a reentrant behavior
in low dimensions $(1+1)< D < (2.065 + 1) $.
}
\end{figure}

So far,
the transverse-field-driven phase transition 
at $H=H_c(\eta)$
with the {\em fixed} anisotropy $\eta$
has been investigated extensively
by means of
the exact-diagonalization \cite{Fey16,Homrighausen17},
series-expansion \cite{Adelhardt20,Fey16}, 
matrix-product-state \cite{Koffel12,Vanderstraeten18}  
and density-matrix-renormalization-group \cite{Frerot17,Zhu18} 
methods.
On the contrary,
little attention has been paid to
the anisotropy-driven criticality,
namely, the multi-criticality
at the $XX$-symmetric point $\eta=0$.
As shown in Fig. \ref{figure2},
for the large-$N$ magnet \cite{Wald15},
the phase boundary $H_c(\eta)$ exhibits a reentrant behavior 
in low dimensions, $(1+1)< D < (2.065+1)$.
Such a feature indicates a counterintuitive picture
that lower symmetry group indices the disorder phase
in the vicinity of the multi-critical point.
On the one hand,
as for the short-range $XY$ magnet,
only the cases
of $D=(2+1)$ \cite{Wald15,Henkel84,Jalal16,Nishiyama19}
and $(1+1)$ \cite{Mukherjee11}
have been considered.
The former shows the multi-criticality
\cite{Riedel69,Pfeuty74}
$H_c(\eta)-H_c(0) \sim |\eta|^{1/\phi}$ with 
the crossover exponent $\phi=1$,
whereas the latter exhibits no singularity
$H_c(\eta)-H_c(0) =0$ at all.
Then, there arises a problem how the crossover exponent
behaves at a transient point $D = 2.5$.
It might be anticipated that $H_c(\eta)$'s slope grows monotonically with
$\phi=1$ retained,
as the dimensionality $D$ increases.
However,
in principle, 
the phase boundary can be curved
convexly, accompanied with a suppressed exponent $\phi<1$ in such a low-dimensionality regime.
The aim of this paper is to explore the multi-criticality
for the transient dimensionality 
by adjusting the decay rate $\sigma$ for the long-range $XY$ magnet (\ref{Hamiltonian}).


As a probe to detect the phase transition
\cite{Quan06,Zanardi06,HQZhou08,Yu09,You11},
we resort to the fidelity
\cite{Uhlmann76,Jozsa94,Peres84,Gorin06}
\begin{equation}
\label{fidelity}
F(H,\Delta H)=  |\langle H | H+\Delta H\rangle |,
\end{equation}
with the ground states,
$|H\rangle$ and $|H+\Delta H\rangle$,
for the proximate interaction parameters,
$H$ and $H+\Delta H$, respectively.
The fidelity 
(\ref{fidelity})
is readily accessible via the exact-diagonalization method,
which admits the ground-state vector $|H\rangle$ explicitly.
Moreover,
the fidelity does not rely on any presumptions
as to the order parameters involved
\cite{Wang15}, 
and it is sensitive to 
generic types of phase transitions such
as the $XX$- ($\eta=0$)
and
$XY$-symmetric ($\eta \ne 0$) cases.
In fairness, 
it has to be mentioned that
the information-theoretical quantifier, the so-called
genuine multipartite entanglement, detects the phase boundary
$H_c(\eta)$ clearly
for rather restricted system sizes, $N \le 20$
\cite{Roy18}.
In this paper,
we treated the cluster
with $N \le 32$ spins,
taking the advantage in that the fidelity 
(\ref{fidelity})
is computationally less demanding,

The rest of this paper is organized as follows.
In Sec. \ref{section2},
the numerical results are presented.
In the last section, we address the summary and discussions.

\section{\label{section2}Numerical results}

In this section, we present the numerical results
for the transverse-field $XY$ chain with the long-range interactions (\ref{Hamiltonian}).
We employed the exact-diagonalization method
for the cluster with $N \le 32$ spins.
Because the exact-diagonalization method yields the ground-state
vector $| H \rangle $ explicitly,
one is able to calculate
the fidelity, namely, the overlap between the proximate parameters,
$F=| \langle H | H+\Delta H\rangle |$ (\ref{fidelity}),
straightforwardly.
Thereby, we evaluated
the fidelity susceptibility 
\cite{Quan06,Zanardi06,HQZhou08,Yu09,You11},
\begin{equation}
	\label{fidelity_susceptibility}
\chi_F = - \frac{1}{N}
	\partial_{\Delta H}^2 F(H,H+\Delta H)|_{\Delta H=0}
	,
\end{equation}
in order to detect the signature for the criticality.
The fidelity susceptibility yields rather reliable estimates for the criticality,
even though the available system size is restricted
\cite{Yu09}.
According to Ref. \cite{Albuquerque10},
the fidelity susceptibility obeys the scaling formula
\begin{equation}
\label{scaling_formula}
\chi_F= N^{\alpha_F/\nu} f\left((H-H_c)N^{1/\nu}\right) ,
\end{equation}
with a certain scaling function $f$.
Here, the indices $\alpha_F$ 
and $\nu$ denote the fidelity-susceptibility and correlation-length ($\xi$)
critical exponents, respectively,
and these exponents 
describe the power-law singularities
of the respective quantities
such as
$\chi_F \sim |H-H_c|^{-\alpha_F}$
and
$\xi \sim |H-H_c|^{-\nu}$.

In fairness, it has to be mentioned that
the fidelity susceptibility
(\ref{fidelity_susceptibility})
was utilized successfully for the analysis of the multi-criticality
in $D=(1+1)$ dimensions \cite{Mukherjee11}.
In this elaborated work, the authors took a direct route toward the
multi-critical point;
here,
the signal of the fidelity susceptibility  splits into sequential subpeaks,
reflecting the intermittent level crossings along $\eta=0$.
Here, we took an indirect route to the multi-critical point
with the properly scaled $\eta$
for each $N$ through resorting to
the crossover-scaling theory 
\cite{Riedel69,Pfeuty74}.
Before commencing detailed crossover-scaling analyses of the multi-criticality,
we demonstrate the performance of the
$\chi_F$-mediated simulation scheme with the fixed anisotropy $\eta$.

\subsection{\label{section2_1}
Fidelity-susceptibility analysis of 
the critical point $H_c$
for the fixed $(\sigma,\eta)=(1,0.5)$}

As a preliminary survey, 
by the agency of the fidelity susceptibility 
(\ref{fidelity_susceptibility}),
we analyze the critical point $H_c$
with the fixed
interaction parameters to
$\sigma=1$ and $\eta=0.5$,
for which an elaborated
series-expansion result \cite{Adelhardt20}
is available.

\begin{figure}
\includegraphics{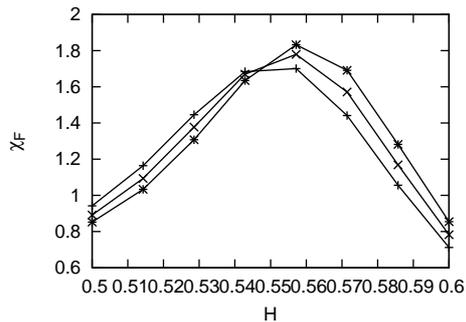}
\caption{
	\label{figure3}
The fidelity susceptibility 
$\chi_F$
(\ref{fidelity_susceptibility})
is plotted for various values of the 
transverse field $H$ 
and the system sizes, 
($+$) $N=28$,
($\times$) $30$, and 
($*$) $32$,
with 
the algebraic decay rate $\sigma=1$
and 
the
$XY$ anisotropy $\eta=0.5$.
The fidelity-susceptibility
peak 
around $H \approx 0.55$
indicates the onset of
the transverse-field-driven phase transition.
}
\end{figure}

In Fig. \ref{figure3},
we present
the
fidelity susceptibility 
$\chi_F$
for various values of the transverse field $H$
and the system sizes,
($+$) $N=28$,
($\times$) $30$ and
($*$) $32$,
with the fixed interaction parameters,
$(\sigma,\eta)=(1,0.5)$.
The fidelity susceptibility exhibits a
pronounced peak around $H\approx 0.55$,
indicating that the transverse-field-driven phase transition
separating the order ($H<H_c$)
and
disorder ($H>H_c$) phases
takes place.


\begin{figure}
\includegraphics{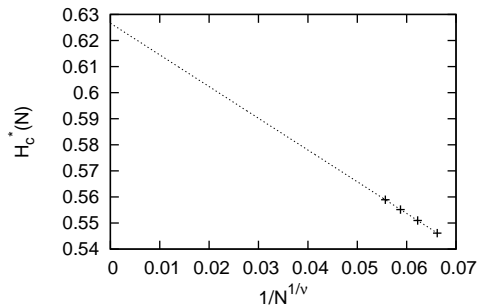}
\caption{
\label{figure4}
The approximate critical point 
$H_c^*(N)$ (\ref{approximate_critical_point})
is plotted for $1/N^{1/\nu}$ with the fixed interaction parameters
$(\sigma,\eta)=(1,0.5)$, and the inverse correlation-length exponent $1/\nu$ 
(\ref{approximate_inverse_correlation_length_exponent}).
The least-squares fit to these data yields 
an estimate
$H_c=0.62665(5)$ in the thermodynamic limit $N\to\infty$.
A possible systematic  error is considered in the text.
This result agrees with the series-expansion result 
(\ref{series_expansion_critical_point}).
}
\end{figure}

Aiming to estimate the 
critical point precisely,
in Fig. \ref{figure4},
we present the approximate
critical point
$H_c^*(N)$
for $1/N^{1/\nu}$ \cite{Albuquerque10}
with the fixed $(\sigma,\eta)=(1,0.5)$.
Here,
the approximate critical point
$H_c^*(N)$
denotes the location of the 
fidelity-susceptibility peak
\begin{equation}
\label{approximate_critical_point}
\partial_H \chi_F(N)|_{H=H_c^*(N)}
=0
,
\end{equation}
for each $N$,
and as mentioned above,
the index $\nu$ describes
correlation-length's singularity,
{\it i.e.},
$\xi \sim |H-H_c|^{-\nu}$.
The abscissa scale
$1/N^{1/\nu}$ 
\cite{Albuquerque10}
comes from this formula;
actually, 
the relation
$H_c^*(N) -H_c \sim N^{1/\nu}$
follows immediately,
because the correlation length
has the same scaling dimension 
as that of the system size \cite{Albuquerque10}, $\xi \sim N $.
The inverse-correlation-length critical
exponent is expressed as
\begin{equation}
\label{approximate_inverse_correlation_length_exponent}
1/\nu = \sigma/2+1/3
 ,
\end{equation}
from the scaling formula 
$\nu=2 \nu_{SR}(D) / \sigma$
($\nu_{SR}(D)$: correlation-length exponent
for the short-range $D$-dimensional counterpart) \cite{Defenu17},
the $\epsilon$-expansion result
$1/\nu_{SR}(D)=2-(4-D)/3$ \cite{Amit05},
and Eq. (\ref{effective_dimensionality}).
This approximate expression $1/\nu$
(\ref{approximate_inverse_correlation_length_exponent}) is not
used
in the crossover-scaling analysis in Sec. \ref{section2_3},
which is the main concern of this paper;
note that
the multi-critical behavior is not identical to  that of
the  aforementioned  transverse-field-driven one.


The least-squares fit to the data
in Fig. \ref{figure4}
yields an estimate
$H_c=
0.62665(5)$
in the thermodynamic limit $N\to\infty$.
Clearly, the estimate may be affected 
by the extrapolation errors.
In order to appreciate a possible systematic error,
we carried out the same extrapolation scheme
with an alternative
$1/\nu[=\sigma/(2-0.037)/0.63] \approx 0.81$ obtained 
via Eq. (32) of Ref. \cite{Defenu17} and the 
3D-Ising result \cite{Deng03}.
Replacing the abscissa scale with this value $1/\nu=0.81$,
we arrived at an alternative one $H_c=0.62877$.
The deviation from the aforementioned estimate
$H_c=0.62665$
appears to
 be
$\approx 2\cdot 10^{-3}$, which seems to dominate the
aforementioned least-squares-fitting error $5\cdot 10^{-5}$;
namely,
the deviation between these independent extrapolation schemes
indicates an appreciable systematic error.
Hence,
considering the former $2 \cdot 10^{-3}$ as the error margin,
we estimate the critical point
as
\begin{equation}
\label{critical_point}
H_c=0.6267(20)
.
\end{equation}
This result appears to 
agree with 
the series-expansion 
result
\cite{Adelhardt20}
\begin{equation}
\label{series_expansion_critical_point}
H_c \approx 0.627
,
\end{equation}
for $\sigma=1$ and $\eta=0.5$.
This value (\ref{series_expansion_critical_point})
was read off by the present author from Fig. 3 of Ref
\cite{Adelhardt20};
here,
the Kac factor 
${\cal N}=4 \zeta(2)$
(\ref{Kac_factor})
has to be multiplied so 
as to 
remedy the energy-scale difference.

A few remarks are in order.
First,
the agreement between 
the sophisticated-series-expansion estimate
$H_c \approx 0.627$ 
[Eq. (\ref{series_expansion_critical_point})], and ours 
$H_c=0.6267(20)$ 
[Eq. (\ref{critical_point})]
confirms the validity of the fidelity-susceptibility-mediated 
simulation scheme.
Second,
as seen from Fig. \ref{figure3},
the back-ground contributions (non-singular part) as to the
fidelity-susceptibility peak appear to be 
rather suppressed.
Actually, as argued in the next section,
the critical exponent of the fidelity susceptibility is
substantially larger than that of the specific heat,
and the fidelity susceptibility exhibits a pronounced
signature for the criticality.
Such a character was reported
by the exact-diagonalization analysis
for the two-dimensional $XXZ$ magnet with the
restricted $N \le 20$ \cite{Yu09}.
Last,
we explain
the reason why only the large system sizes
$N \approx 30$
were treated in the extrapolation analysis as in Fig. \ref{figure4}.
Because we are considering the crossover critical phenomenon,
the series of finite-size data exhibit two types of
scaling behaviors,
such as
the Ising- and $XX$-type singularities,
as $N$ increases.
In our preliminary survey,
the large system sizes $N \approx 30$
were found to capture the desired scaling behavior
coherently,
at least, for the critical domain undertaken in this simulation study.
Because of this reason, the extrapolation scheme in Fig. \ref{figure4}
cannot be straightforwardly replaced with the
sophisticated extrapolation sequences.

\subsection{\label{section2_2}
Scaling analysis of the fidelity susceptibility
for  $(\sigma,\eta)=(1,0.5)$}

In this section,
we investigate the critical behavior of the fidelity susceptibility,
following the analyses in the preceding section.
For that purpose,
we 
rely on
the scaling theory (\ref{scaling_formula})
for the fidelity susceptibility 
developed in Ref. \cite{Albuquerque10};
afterward,
this scaling theory is extended 
in order to investigate the multi-criticality
\cite{Riedel69,Pfeuty74}
around $\eta=0$.
According to the scaling argument
\cite{Albuquerque10}
the scaling dimension $\alpha_F/\nu$ 
of the fidelity susceptibility
satisfies the relation
\begin{equation}
\label{scaling_relation}
\alpha_F/\nu=\alpha/\nu +z
,
\end{equation}
with 
the dynamical critical exponent
\cite{Defenu17}
\begin{equation}
\label{dynamical_critical_exponent}
z=\sigma/2,
\end{equation}
and
the specific-heat critical exponent $\alpha$;
namely, the specific heat exhibits the singularity
as
$C\sim|H-H_c|^{ - \alpha}$.
Notably enough, the scaling dimension $\alpha_F/\nu$
of the fidelity susceptibility is larger
than that of the specific-heat $\alpha / \nu$,
indicating that the former
should exhibit a pronounced 
singularity, as compared to the latter.
As a consequence,
we arrive at the expression
\begin{equation}
\label{approximate_fidelity_susceptibility_scaling_dimension}
\alpha_F/\nu=\sigma -1/3
 ,
\end{equation}
from
the hyper-scaling relation 
$\alpha=2-(1+z)\nu$ \cite{Albuquerque10},
and Eq.
(\ref{approximate_inverse_correlation_length_exponent}),
(\ref{scaling_relation}),
and 
(\ref{dynamical_critical_exponent}).
The critical indices associated with the above scaling formula 
(\ref{scaling_formula})
are all fixed, and now, we are able to carry out the scaling analysis
of the fidelity susceptibility without 
any adjustable parameters.

\begin{figure}
\includegraphics{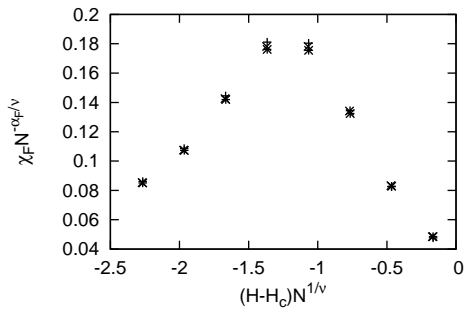}
\caption{\label{figure5}
The scaling plot of the fidelity susceptibility,
$(H-H_c)N^{1/\nu}$-$N^{-\alpha_F/\nu} \chi_F$,
is presented for various system sizes, 
($+$) $N=28$,
($\times$) $30$, and
($*$) $32$,
with $(\sigma,\eta)=(1,0.5)$.
Here, the scaling parameters, critical point,
inverse correlation-length
exponent
and $\chi_F$'s scaling dimension, are set to
$H_c=0.6267$ (\ref{critical_point}),
$1/\nu=5/6$ (\ref{approximate_inverse_correlation_length_exponent}),
and
$\alpha_F/\nu=2/3$ (\ref{approximate_fidelity_susceptibility_scaling_dimension}), respectively.
The fidelity-susceptibility data obey the
scaling theory (\ref{scaling_formula})
rather satisfactorily.
}
\end{figure}

In Fig. \ref{figure5},
we present the scaling plot,
$(H-H_c)N^{1/\nu}$-$N^{-\alpha_F/\nu}\chi_F$,
for various system sizes, 
($+$) $N=28$,
($\times$) $30$, and
($*$) $32$,
with the fixed $\sigma=1$ and $\eta = 0.5$.
Here,
the scaling parameters,
$H_c$,
$1/\nu$
and
$\alpha_F/\nu$,
are given by 
Eq. (\ref{critical_point}),
(\ref{approximate_inverse_correlation_length_exponent}),
and
(\ref{approximate_fidelity_susceptibility_scaling_dimension}), respectively.
We observe that the scaled data collapse into a
scaling curve satisfactorily,
confirming the validity of the analysis in Sec. \ref{section2_1}
as well as the scaling argument \cite{Defenu17}
introduced above.


A number of remarks are in order.
First, no 
adjustable parameter was incorporated in the scaling analysis of
Fig. \ref{figure5}.
Actually, the scaling parameters, 
$H_c$, 
$\nu$, and 
$\alpha_F/\nu$, 
were fixed in prior to undertaking the scaling analysis.
Last, the scaling data, Fig. \ref{figure5}, appear to be less influenced
by the finite-size artifact,
indicating that the simulation data
already enter into the scaling regime.
It is a benefit of the fidelity susceptibility that 
it is not influenced by corrections to scaling very severely \cite{Yu09}.
Encouraged by this finding, we explore the multi-critical behavior 
via $\chi_F$ in the next section.

\subsection{\label{section2_3}
Crossover-scaling analysis of the fidelity susceptibility
for $\sigma=4/3$ ($D=2.5$) around the $XX$-symmetric point}

In this section, we investigate the 
crossover-scaling
(multi-critical) behavior
around the $XX$-symmetric point $\eta=0$
via the fidelity susceptibility.
Here,
we set the decay rate to $\sigma=4/3$,
which corresponds to $D=2.5$ according to 
Eq.
(\ref{effective_dimensionality}).
As mentioned
in Introduction,
the $D=2$ \cite{Mukherjee11}
and $3$ 
\cite{Wald15,Henkel84,Jalal16,Nishiyama19}
cases have been studied,
and the transient behavior in between remains unclear.

For that purpose, we incorporate a yet another parameter 
$\eta$ accompanied
with the crossover exponent $\phi$.
Then, the aforementioned
expression (\ref{scaling_formula})
is extended 
to the crossover-scaling formula
\cite{Riedel69,Pfeuty74}
\begin{equation}
	\label{crossover_scaling_formula}
\chi_F=N^{\dot{\alpha}_F/\dot{\nu}} 
g\left((H-H_c(\eta))N^{1/\dot{\nu}},
	\eta N^{\phi/\dot{\nu}}\right)
,
\end{equation}
with a certain scaling function $g$.
Here,
the symbol $H_c(\eta)$ denotes the
critical point for each $\eta$,
and
the indices $\dot{\nu}$ and $\dot{\alpha}_F$
denote the correlation-length
and fidelity-susceptibility critical exponents,
respectively,
right at the multi-critical point $\eta=0$;
namely, respective singularities are given by
$\xi \sim |H-H_c(0)|^{-\dot{\nu}}$
and 
$\chi_F \sim |H-H_c(0)|^{-\dot{\alpha}_F}$.
The former relation together with $N\sim\xi$ 
immediately yields \cite{Riedel69,Pfeuty74}
\begin{equation}
H_c(\eta) -H_c(0) \sim |\eta|^{1/\phi}
,
\end{equation}
because the second argument of the crossover-scaling formula
(\ref{crossover_scaling_formula}),
$\eta N^{\phi/\dot{\nu}}$, should be dimensionless (scale-invariant).
Hence, the crossover exponent $\phi$ governs the power-law singularity
of the phase boundary.
As in Eq. (\ref{scaling_relation}),
these critical indices satisfy
the scaling relation \cite{Albuquerque10}
\begin{equation}
\label{crossover_scaling_relation}
\dot{\alpha}_F/\dot{\nu}=\dot{\alpha}/\dot{\nu}+\dot{z},
\end{equation}
with the specific-heat and dynamical critical exponents,
$\dot{\alpha}$ and 
$\dot{z}$, respectively, at the multi-critical point.

Before commencing the crossover-scaling analyses of $\chi_F$,
we fix the set of the multi-critical indices appearing in Eq. 
(\ref{crossover_scaling_formula}).
At the $XX$-symmetric point, the critical indices 
for the transverse-field-driven phase transition
were determined \cite{Zapf14}
as
$\dot{\alpha}=1/2$,
$\dot{z}=2z$ 
and
\begin{equation}
\label{multicritical_inverse_correlation_length_exponent}
1/\dot{\nu}=\sigma
  . 
\end{equation}
This index (\ref{multicritical_inverse_correlation_length_exponent})
is taken from 
Eq. (19) of Ref. \cite{Defenu17}, as it means
the mean-field value \cite{Zapf14}.
Hence, from Eq. (\ref{dynamical_critical_exponent}) and (\ref{crossover_scaling_relation}),
we arrive at 
\begin{equation}
	\label{crossover_scaling_dimension}
\dot{\alpha}_F/\dot{\nu}= 3\sigma/2
.
\end{equation}
The above argument completes the prerequisite for the crossover-scaling analysis.
The index $\phi$ has to be determined so as to attain
a good data collapse of the crossover-scaling plot,
based on the formula (\ref{crossover_scaling_formula}).

\begin{figure}
\includegraphics{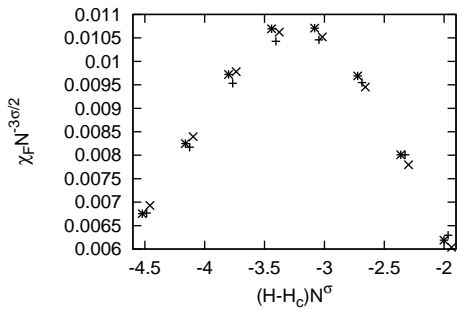}
\caption{
	\label{figure6}
The crossover-scaling plot
	of the fidelity susceptibility,
	$(H-H_c(\eta))N^{1/\dot{\nu}}$-$N^{-\dot{\alpha}_F/\dot{\nu}} \chi_F$,
is presented for various system sizes,
($+$) $N=28$,
($\times$) $30$, and
($*$) $32$,
with the fixed decay rate $\sigma=4/3$ corresponding to $D=2.5$ 
[Eq. (\ref{effective_dimensionality})].
Here,
	the second argument of the crossover-scaling formula (\ref{crossover_scaling_formula})
is fixed to $\eta N^{\phi/\dot{\nu}}=15.2$ with the optimal crossover exponent
$\phi=1$, and
the other multi-critical indices are set to
$1/\dot{\nu}=\sigma$ (\ref{multicritical_inverse_correlation_length_exponent})
and $\dot{\alpha}_F/\dot{\nu}=3\sigma/2$ (\ref{crossover_scaling_dimension}).
The crossover-scaled data collapse into the scaling curve
rather satisfactorily
under the optimal setting $\phi=1$.
}
\end{figure}

In Fig. \ref{figure6},
we present the crossover-scaling plot,
$(H-H_c(\eta))N^{1/\dot{\nu}}$-$N^{-\dot{\alpha}_F/\dot{\nu}}\chi_F$,
for various system sizes,
($+$) $N=28$,
($\times$) $30$, and 
and
($*$) $32$,
with $\sigma=4/3$, 
$1/\dot{\nu}=\sigma$ (\ref{multicritical_inverse_correlation_length_exponent}),
and
$\dot{\alpha}_F/\dot{\nu}=3\sigma/2$ (\ref{crossover_scaling_dimension}).
Here, the second argument of the crossover-scaling formula
(\ref{crossover_scaling_formula})
is fixed to $\eta N^{\phi/\dot{\nu}}=15.2(\approx 0.15\cdot 32^{\phi/\dot{\nu}})$ 
under the optimal crossover exponent $\phi=1$,
and the critical point $H_c$ was determined via the same scheme as
that of Sec. \ref{section2_1}.
From Fig. \ref{figure6},
we see that 
the crossover-scaled data
fall into a scaling curve satisfactorily,
indicating that the choice $\phi=1$ should be a feasible one.


\begin{figure}
\includegraphics{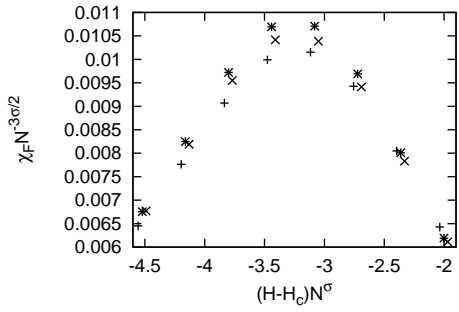}
\caption{
	\label{figure7}
The crossover-scaling plot
	of the fidelity susceptibility,
	$(H-H_c(\eta))N^{1/\dot{\nu}}$-$N^{-\dot{\alpha}_F/\dot{\nu}} \chi_F$,
is presented for various system sizes,
($+$) $N=28$,
($\times$) $30$, and
($*$) $32$,
with the fixed decay rate $\sigma=4/3$.
Here,
	the second argument of the crossover-scaling formula (\ref{crossover_scaling_formula})
is fixed to $\eta N^{\phi/\dot{\nu}}=30.5$ with a slightly large crossover exponent $\phi=1.15$;
	the other multi-critical indices, $\dot{\nu}$ and $\dot{\alpha}_F/\dot{\nu}$,
	are the same as those of Fig. \ref{figure6}.
	For such a large value of $\phi=1.15$,
	the hilltop data get scattered.
}
\end{figure}

\begin{figure}
\includegraphics{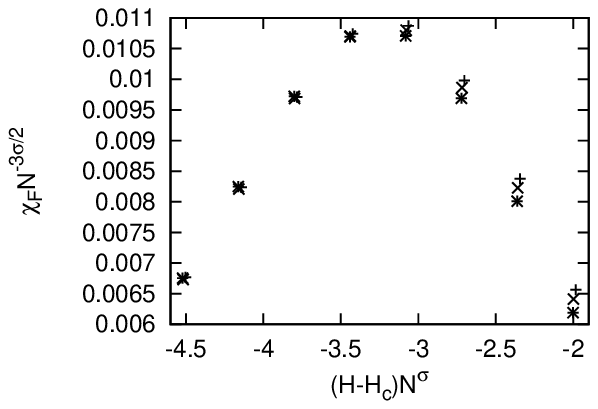}
\caption{
	\label{figure8}
The crossover-scaling plot
	of the fidelity susceptibility,
	$(H-H_c(\eta))N^{1/\dot{\nu}}$-$N^{-\dot{\alpha}_F/\dot{\nu}} \chi_F$,
is presented for various system sizes,
($+$) $N=28$,
($\times$) $30$, and
($*$) $32$,
with the fixed decay rate $\sigma=4/3$.
Here,
	the second argument of the crossover-scaling formula (\ref{crossover_scaling_formula})
is fixed to $\eta N^{\phi/\dot{\nu}}=7.62$ with a 
slightly small crossover exponent $\phi=0.85$;
	the other multi-critical indices, $\dot{\nu}$ and $\dot{\alpha}_F/\dot{\nu}$,
	are the same as those of Fig. \ref{figure6}.
	For such a small value of $\phi=0.85$,
	the right-side-slope data start to split off.
}
\end{figure}

Setting the crossover exponent to a slightly large value $\phi=1.15$,
in Fig. \ref{figure7},
we present the crossover-scaling plot,
$(H-H_c(\eta))N^{1/\dot{\nu}}$-$N^{-\dot{\alpha}_F/\dot{\nu}}\chi_F$,
for various system sizes $N=28,30,32$; 
the symbols and the critical indices,
$1/\dot{\nu}$ and 
$\dot{\alpha}_F/\dot{\nu}$, are the same as those of 
Fig. \ref{figure6}.
Here, the second argument of the crossover-scaling formula (\ref{crossover_scaling_formula})
is set to 
$\eta N^{\phi/\dot{\nu}}=30.5$
with $\phi=1.15$.
For such a large vale of $\phi=1.15$,
the hilltop data get scattered, as compared to those of Fig. \ref{figure6}.
Likewise,
in Fig. \ref{figure8},
for a small value of $\phi=0.85$,
we present the crossover-scaling plot,
$(H-H_c(\eta))N^{1/\dot{\nu}}$-$N^{-\dot{\alpha}_F/\dot{\nu}}\chi_F$,
for various system sizes $N=28,30,32$; 
the symbols and
the critical indices, 
$1/\dot{\nu}$ and 
$\dot{\alpha}_F/\dot{\nu}$, are the same as those of 
Fig. \ref{figure6}.
Here, the second argument of the crossover-scaling formula (\ref{crossover_scaling_formula})
is set to 
$\eta N^{\phi/\dot{\nu}}=7.62$
with $\phi=0.85$.
For such a small value of $\phi=0.85$,
the right-side-slope data seem to split off.
Considering that the cases, Fig. \ref{figure7} and \ref{figure8},
set the upper and lower bounds, respectively, for 
$\phi$,
we conclude that the crossover exponent lies within
\begin{equation}
\label{crossover_exponent}
\phi=1.00(15)
.
\end{equation}


This result
(\ref{crossover_exponent}) indicates that the phase boundary
for $\sigma=4/3$ ($D=2.5$) rises up linearly,
$H_c(\eta) \sim |\eta|$,
around the multi-critical point $\eta=0$;
see Fig. \ref{figure1}.
Namely, the multi-criticality is identical to that of $D=3$ 
\cite{Wald15,Henkel84,Jalal16,Nishiyama19}.
Hence, it is suggested that the linearity is robust against the dimensionality $D$,
and merely, the slope grows up monotonically, as the dimensionality $D$ increases
from $D=2$.
In other words,
no  exotic character such as the reentrant behavior predicted
by the large-$N$ theory occurs, at least, for the $XY$ magnet.

A number of remarks are in order.
First,
underlying physics behind the crossover-scaling plot,
Fig. \ref{figure6},
differs significantly from that of the fixed-$\eta$ scaling plot,
Fig. \ref{figure5}.
Actually, the former scaling dimension,
$\dot{\alpha}_F/\dot{\nu}=3\sigma/2$ (\ref{crossover_scaling_dimension}),
is larger than that of the latter, $\alpha_F/\nu=\sigma - 1/3$
(\ref{approximate_fidelity_susceptibility_scaling_dimension}).
Hence, the data collapse in Fig. \ref{figure6} is by no means accidental, 
and accordingly,
the crossover exponent $\phi$ has to be adjusted rather carefully.
Second, 
we stress that the critical indices other than $\phi$ were fixed
in prior to performing the crossover-scaling analyses.
Last,
in the $\chi_F$-mediated analysis,
no presumptions as to the order parameters are made.
Because in our study, the crossover between the $XX$- and $XY$-symmetric cases is 
concerned, it is significant that the quantifier is sensitive to both
order parameters in a systematic manner.

\section{\label{section3}Summary and discussions}

We investigated
the transverse-field $XY$ chain with the long-range interactions
(\ref{Hamiltonian})
by means of the exact-diagonalization method.
Because the method allows us to access directly to the ground state,
one does not have to care about the anisotropy
between the
real-space and imaginary-time directions rendered by
the dynamical critical exponent $z(\ne 1)$ (\ref{dynamical_critical_exponent}).
As a preliminary survey, setting the decay rate and the $XY$ anisotropy 
to $\sigma=1$ and $\eta=0.5$, respectively,
we analyzed the transverse-field-driven phase transition
by the agency of the fidelity susceptibility (\ref{fidelity_susceptibility}).
Our result
$H_c=0.6267(20)$ [Eq. (\ref{critical_point})] 
agrees
with the preceeding 
series-expansion
result $H_c \approx 0.627$ 
[Eq. (\ref{series_expansion_critical_point})],
confirming the validity of our simulation scheme.
Actually,
as the scaling relation (\ref{scaling_relation}) indicates,
the scaling dimension of the 
fidelity susceptibility, $\alpha_F/\nu$,
is larger than that of the specific heat,
$\alpha/\nu$,
and 
the former 
exhibits a pronounced peak,
as shown in  Fig. \ref{figure3}.
We then 
turn to the analysis of the multi-criticality around the 
$XX$-symmetric point $\eta=0$ under the setting
$D=2.5$ ($\sigma=4/3$).
Scaling the
the $XY$-anisotropy parameter $\eta$ properly,
we cast
the fidelity-susceptibility data into the crossover-scaling formula 
(\ref{crossover_scaling_formula}).
The crossover-scaled data fall into the scaling curve
satisfactorily
under the setting, $\phi=1.00(15)$ [Eq. (\ref{crossover_exponent})].
The result indicates that the phase boundary $H_c(\eta)$ 
rises up linearly, $\sim |\eta|$,
around the multi-critical point for $D=2.5$.
Such a character is 
identical to that of
the $D=(2+1)$ case,
suggesting that $H_c(\eta)$'s
linearity is retained in low dimensions,
at least, for the $XY$ magnet.

Then, there arises
a problem whether the crossover exponent
is influenced by the extention of the internal symmetry group.
Actually, as for the large-$N$ magnet,
the phase boundary should exhibit
a reentrant behavior 
within $2< D < 3.065\dots$.
We conjecture that for sufficiently large internal-symmetry group,
the phase boundary gets curved convexly, $\phi \approx 0.5$, 
for $D \approx 2.5$,
and above this threshold, the reentrant behavior eventually sets in.
The SU$(4)$ magnet \cite{Itoi00}
subjected to the transverse field would be a promising candidate to examine this scenario,
and this problem is left for the future study.

\section*{Acknowledgment}
This work was supported by a Grant-in-Aid
for Scientific Research (C)
from Japan Society for the Promotion of Science
(Grant No. 
20K03767).

{\bf Author contribution statement}

The presented idea was conceived by Y.N.
He also performed the computer simulations, analyzed 
the data, and wrote the manuscript.

%

%
%



\begin{thebibliography}{}
%



\bibitem{Maziero10}
J. Maziero, H. C. Guzman, L. C. C\'{e}leri, M. S. Sarandy, and R. M. Serra,
Phys. Rev. A {\bf 82} (2010) 012106.  
\bibitem{Sun14}
Z.-Y. Sun, Y.-Y. Wu, J. Xu, H.-L. Huang, B.-F. Zhan, B. Wang, and C.-B. Duanpra,
Phys. Rev. A {\bf 89} (2014) 022101.

\bibitem{Karpat14}
G. Karpat, B. \c{C}akmak, and F. F. Fanchini,
Phys. Rev. B {\bf 90} (2014) 104431. 




\bibitem{Luo18}
Q. Luo, J. Zhao, and X. Wang,
Phys. Rev. E {\bf 98} (2018) 022106.


\bibitem{Steane98}A. Steane, Rep. Prog. Phys. {\bf 61} (1998) 117.
\bibitem{Bennett00}C.H. Bennett and D.P. DiVincenzo, Nature {\bf 404} (2000) 247.



\bibitem{Adelhardt20}
P. Adelhardt, J. A. Koziol, A. Schellenberger, and K. P. Schmidt,
Phys. Rev. B {\bf 102} (2020) 174424.




\bibitem{Defenu17}
N. Defenu, A. Trombettoni, and S. Ruffo
Phys. Rev. B {\bf 96} (2017) 104432.





\bibitem{Dutta01}
A. Dutta and J. K. Bhattacharjee,
Phys. Rev. B {\bf 64} (2001) 184106.
%

\bibitem{Koffel12}
T. Koffel, M. Lewenstein, and L. Tagliacozzo,
Phys. Rev. Lett. {\bf 109} (2012) 267203.  

\bibitem{Campana10}
L. S.  Campana, L. De Cesare, U. Esposito, M. T.  Mercaldo,
and I. Rabuffo,
Phys. Rev. B {\bf 82} (2010) 024409.
\bibitem{Gong16}
Z.-X. Gong, M. F.  Maghrebi, A.  Hu, M. Foss-Feig, P. Richerme,
C. Monroe, and A. V. Gorshkov,
Phys. Rev. B {\bf 93} (2016) 205115.


\bibitem{Fey16}
S. Fey and K. P. Schmidt, Phys. Rev. B {\bf 94} (2016) 075156.

%
\bibitem{Maghrebi17}
M. F. Maghrebi, Z.-X. Gong, and A. V. Gorshkov,
Phys. Rev. Lett. {\bf 119} (2017) 023001.

\bibitem{Frerot17}
I. Fr\'erot, P. Naldest, and T. Roscilde, 
Phys. Rev. B {\bf 95} (2017) 245111.

\bibitem{Roy18}
S. S. Roy and H. S. Dhar,
Phys. Rev. A {\bf 99} (2019) 062318.


\bibitem{Puebla19}
R. Puebla, O. Marty, and M. B. Plenio, Phys. Rev A {\bf 100} (2019) 032115. 



\bibitem{Fisher72}
M. E. Fisher, S.-k. Ma, and B. G. Nickel,
Phys. Rev. Lett. {\bf 29} (1972) 917.
\bibitem{Sak73}
J. Sak, Phys. Rev. B {\bf 8} (1973) 281. 


\bibitem{Gori17}
G. Gori, M. Michelangeli, N. Defenu, and A. Trombettoni
Phys. Rev. E {\bf 96} (2017) 012108. 
\bibitem{Angelini14}
M. C. Angelini, G. Parisi, and F. Ricchi-Tersenghi,
Phys. Rev. E {\bf 89} (2014) 062120. 
\bibitem{Joyce66}
J. S. Joyce, Phys. Rev. {\bf 146} (1966) 349. 



\bibitem{Wald15}
S. Wald and M. Henkel, J. Stat. Mech.: Theory and Experiment
(2015) P07006. 

\bibitem{Henkel84}
M. Henkel, J. Phys. A: Mathematical and Theoretical {\bf 17} (1984) L795.

\bibitem{Jalal16}
S. Jalal, R. Khare, and S. Lal, arXiv:1610.09845.

\bibitem{Nishiyama19}
Y. Nishiyama,
Eur. Phys. J. B {\bf 92} (2019) 167.


\bibitem{Mukherjee11}
V. Mukherjee, A. Polkovnikov, and A. Dutta,
Phys. Rev. B {\bf 83} (2011) 075118. 




\bibitem{Homrighausen17}
I. Homrighausen, N. O. Abeling, V. Zauner-Stauber, J. C. Halimeh,
Phys. Rev. B {\bf 96} (2017) 104436.

\bibitem{Vanderstraeten18}
L. Vanderstraeten, M. Van Damme, H. P. B\"uchler, F. Verstraete,
Phys. Rev. Lett. {\bf 121} (2018) 090603.




\bibitem{Goll18}
R. Goll and P. Kopietz, 
Phys. Rev. E {\bf 98} (2018) 022135.



\bibitem{Luijten02}
E. Luijten and H. W. J. Bl\"ote,
Phys. Rev. Lett. {\bf 89} (2002) 025703. 
\bibitem{Brezin14} 
E. Brezin, G. Parisi, and F. Ricci-Tersenghi,
J. Stat. Phys. {\bf 157} (2014) 855. 
\bibitem{Defenu15} 
N. Defenu, A. Trombettoni, and A. Codello,
Phys Rev. E {\bf 92} (2015) 052113. 




\bibitem{Zhu18}
Z. Zhu, G. Sun, W.-L. You, and D.-N. Shi,  
Phys. Rev. A {\bf 98} (2018) 023607.






\bibitem{Riedel69}E.K. Riedel and F. Wegner, Z. Phys. {\bf 225} (1969) 195.
\bibitem{Pfeuty74}P. Pfeuty, D. Jasnow, and M. E. Fisher,
Phys. Rev. B {\bf 10} (1974) 2088.







\bibitem{Quan06}
H. T. Quan, Z. Song, X. F. Liu, P. Zanardi, and C. P. Sun, 
Phys. Rev. Lett. {\bf 96} (2006) 140604.
%
\bibitem{Zanardi06}
P. Zanardi and N. Paunkovi\'c,
Phys. Rev. E {\bf 74} (2006) 031123.





\bibitem{HQZhou08}
H.-Q. Zhou, and J. P. Barjaktarevi\~c,
J. Phys. A: Math. Theor. {\bf 41} (2008) 412001.




\bibitem{Yu09}
W.-C. Yu, H.-M. Kwok, J. Cao, and S.-J. Gu,
Phys. Rev. E {\bf 80} (2009) 021108.






\bibitem{You11}
W.-L. You and Y.-L. Dong,  
Phys. Rev. B {\bf 84} (2011) 174426.






\bibitem{Uhlmann76}
A. Uhlmann, Rep. Math. Phys. {\bf 9} (1976) 273.
\bibitem{Jozsa94}
R. Jozsa, J. Mod. Opt. {\bf 41} (1994) 2315.

\bibitem{Peres84}
A. Peres, Phys. Rev. A {\bf 30} (1984) 1610.
\bibitem{Gorin06}
T. Gorin, T. Prosen, T. H. Seligman, and M. \v{Z}nidari\v{c}, 
Phys. Rep. {\bf 435} (2006) 33.




\bibitem{Wang15}
L. Wang, Y.-H. Liu, J. Imri\v{s}ka, P. N. Ma, and M. Troyer,
Phys. Rev. X {\bf 5} (2015) 031007.


\bibitem{Albuquerque10}  
A. F. Albuquerque, F. Alet, C. Sire, and S. Capponi,
Phys. Rev. B {\bf 81} (2010) 064418.


\bibitem{Amit05}
D.J. Amit and V. Mart\'in-Mayor, 
{\it Field Theory, the renormalization group, and critical phenomena},
World Scientific 2005


\bibitem{Deng03}
Y. Deng and H.W.J. Bl\"ote, Phys. Rev. E {\bf 68} (2003) 036125.


\bibitem{Zapf14}
V. Zapf, M. Jaime, and C. D. Batista,
Rev. Mod. Phys. {\bf 86} (2014) 563.


\bibitem{Itoi00}
C. Itoi, S. Qin, and I. Affleck,
Phys. Rev. B {\bf 61} (2000)  6747.






\end{thebibliography}


\end{document}